\documentclass[aps,showpacs,prd,twocolumn,superscriptaddress,floatfix]{revtex4}
\usepackage{amsmath, amsfonts, hyperref}
\usepackage{color}
\usepackage{graphicx}
\newcommand{\beq}{\begin{equation}}
\newcommand{\eeq}{\end{equation}}
\newcommand{\beqn}{\begin{eqnarray}}
\newcommand{\eeqn}{\end{eqnarray}}

\def\vhat{\hat v}

\newif\ifdraft
\drafttrue % or \draftfalse
\def\Einf{E_{\infty}}
\def\phat{\hat p}
\def\ninf{n_{\infty}}
\def\vinf{v_{\infty}}
\def\vhat{\hat v}
\def\vhatinf{\hat{v}_{\infty}}
\def\rhoinf{\rho_{\infty}}

\begin{document}

\title
{Spikes and accretion of unbound, collisionless matter around black holes}
\date{\today}
\author{Stuart~L. Shapiro}
\affiliation{Departments of Physics and Astronomy, University of Illinois at Urbana-Champaign, Urbana, IL 61801, USA}
%\altaffiliation{Also NCSA, University of
%  Illinois at Urbana-Champaign, Urbana, IL 61801, USA}

\begin{abstract}
We consider the steady-state density and velocity dispersion profiles of collisionless matter 
around a Schwarzschild black hole (BH) and its associated rate of accretion onto the BH. 
We treat matter, which could be stars or dark matter particles, whose orbits 
are {\it unbound} to the BH, but still governed by its gravitational field. 
We consider two opposite spatial geometries for the matter distributions: an infinite,
3D cluster and a 2D razor-thin disk, both with zero net angular momentum. 
We demonstrate that the results depend critically 
on the adopted geometry, even in the absence of angular momentum.
We adopt a simple monoenergetic, isotropic,  phase-space distribution function for the matter as a 
convenient example to illustrate this dependence. The effect of breaking strict isotropy by 
incorporating an unreplenished loss cone due to BH capture of low-angular momentum matter 
is also considered. Calculations are all analytic and performed in full general relativity, though 
key results are also evaluated in the Newtonian limit. We consider one application to show
that the rate of BH accretion from an ambient cluster can be significantly less than that from 
a thin disk to which it may collapse, although both rates are
considerably smaller than Bondi accretion for a (collisional) fluid with a similar asymptotic
particle density and velocity dispersion (i.e., sound speed).

\end{abstract}

\maketitle

\section{Introduction} 

A black hole (BH), and especially a supermassive black hole (SMBH), 
typically will steepen the density profile of stars and/or dark matter (DM) within 
the hole’s sphere of influence, i.e., within a radius 
$r_h \approx GM /v_{\infty}^2$. Here, $M$ is the mass of the hole and 
$v_{\infty}$ is the velocity dispersion in the cluster or galaxy core outside $r_h$.
The density profile of this spike depends on the 
properties of the matter, as well as the formation and evolution 
history of the BH. Most of the treatments to date
focus on matter {\it bound} to the BH and hence moving on orbits that are confined inside $r_h$.
In some analyses the matter is treated as completely collisionless, 
as was assumed for stars in 
typical galaxies in Refs~\cite{Pee72a,You77}, and for cold DM in galaxies and halos in 
Ref~\cite{GonS99}, where in both of these cases the orbiting species were assumed bound to 
adiabatically growing BHs. 
In other cases the matter is assumed to be weakly collisional and slowly evolving on long 
relaxation timescales. Such relaxation may be  
governed by distant, cumulative, two-body, gravitational encounters (Coulomb scattering) 
for stars bound to a central, massive BH in a globular clusters or to a SMBH in a dense galaxy 
core~\cite{BahW76,FraR76,LigS77,ShaM78,CohK78}. For collisionless DM bound to a SMBH in a 
galaxy core, relaxation may occur due to 
gravitational encounters with ambient stars~\cite{GneP04,Mer04,ShaH22}, or, 
in the case of self-interacting DM (i.e. SIDM) to 
particle self-interactions~\cite{ShaP14,Sha18,DelK23}, and/or annihilations~\cite{Vas7,ShaS16}.
In all of these instances the orbiting matter forms power-law 
density spikes inside $r_h$.

Here we consider 
collisionless matter, either dark matter or stars (both of which we shall refer to as
``particles") that move on orbits {\it unbound} to the BH. These unbound orbits can take particles
infinitely far away from the BH, but particles that plunge inside $r_h$ on such orbits 
also may, in some instances, generate a density spike around the BH. 
Those particles traveling inward with sufficiently
low angular momentum about the BH are eventually captured.
By collisionless here we mean that the particles interact solely with the 
gravitational field of the central BH and
that any scattering due to, or collisions with, their neighbors, themselves or other 
intruders remain generally unimportant over the age of the system.
We take the BH to be a nonspinning Schwarzschild BH with a mass $M$ fixed in time and we determine 
the density profile and accretion rate of the unbound ambient matter. Collsionless particles move
on geodesic orbits about the BH and for this analysis we assume their self-gravity is unimportant. 
We take all of the particles to have the same mass and to describe them we 
adopt a simple, monoenergetic, phase-space distribution function of the form $f=f(E)$, 
where $E$ is the conserved ``energy at infinity" per unit mass of a particle. 
Such a distribution function 
has been adopted in several previous investigations of unbound 
particles about a BH~\cite{ZelN71,ShaT83,ShaT85,GamGDNS21} 
and, though idealized, it is sufficient to illustrate our main conclusions. 
Any distribution function that depends solely on conserved integrals of the 
motion, such as $E$, automatically satisfies the time-independent, collisionless Boltzmann (Vlasov) equation. 
Hence the profiles and accretion rates we determine from our adopted distribution function 
yield steady-state solutions. 

Given our adopted distribution function 
we compare two opposite spatial geometries for the unbound matter  -- 
infinite 3-D clusters and 2-D razor-thin disks. The adopted disks, though extreme, 
serve to highlight the differences between extended, spherical-like vs. thin disk-like density 
distributions for unbound, collisionless particles orbiting a central BH and 
their associated rates of accretion.  
%Such disks have also been treated using 
%a different formalism in 
%~Ref~\cite{CieMO22} for unbound particles with a Maxwell-J\"unter distribution 
%around a Kerr BH. 
We note that collisionless dark matter in the early universe,
and even the first generation stars, may in fact form thin sheets 
or ``pancakes"~\cite{Zel70,KulVBL22}, 
so our extreme disks may mimic some of their features, should massive BHs reside within them. 
In both our cluster and disk cases  we analyze the case where the orbital velocities 
are isotropic at every point. We then consider the anisotropic velocity distribution 
associated with an unreplenished loss cone that can arise from BH capture of particles with sufficiently low angular momentum.  In all cases the net angular momentum of the orbiting matter is assumed to be zero, so we are only examining
the effect of spatial geometry on the profiles and accretion rates.

Infinite, unbound collisionless clusters around Schwarzschild BHs 
have  been previously investigated with a different approach 
in Refs~\cite{RioS17a,RioS17b}, who adopted
a Maxwell-J\"unter phase-space distribution function as an application. 
A treatment involving thin (equatorial plane) disks around Kerr black holes 
was recently provided by Ref~\cite{CieMO22}, again with a Maxwell-J\"unter
distribution function and a different approach from
the one adopted here. We compare the accretion rates found by these studies
with the ones calculated here.
 
The significance of determining the density profiles and accretion rates of stars and DM near BHs is that
it provides clues as to the nature, formation history and evolution of the systems in which they are found.
The presence of a sufficiently steep density spike not only conveys the presence of a massive, central BH 
but, in the case of DM, may lead to an observable excess of gamma rays or other form of radiation, if particle
annihilation occurs at a sufficient rate in the innermost regions where the density is 
highest~[see \cite{DayFHLPRS16,ChoZMS22,LeaETAL22,GonS99,FieSS14,WanBVW15,ChiSS20} and references therein].   
These applications motivate our adopting a simple illustration to point out the importance of 
unbound collisionless particles and their global spatial geometry in determining the existence 
of a stellar, or DM, spike around a BH and their consumption rate by the BH.

In Section II we consider an infinite,
unbound, monoenergetic, 3D  cluster of collisionless particles that orbit a Schwarzschild BH.
We derive the particle density and velocity profiles and the associated particle 
accretion rate onto the BH. 
We consider both an isotropic velocity distribution and one with an unreplenished loss cone.
The derivations and quoted quantities are analytic and fully general relativistic, 
but the final expressions are also quoted
for the slow-velocity, weak-field (i.e. Newtonian) limit. In Section III we
repeat the analysis in Section I, but now for particles confined to a 2D razor-thin disk.
In Section IV we compare the accretion rates found for the clusters and disks. In
the Appendix we provide two alternative derivations for the cluster accretion rate, all
arriving at the same answer. We adopt geometric units throughout, setting
$G=1=c$, unless otherwise noted.

\section{Unbound, Monoenergetic Cluster}

\subsection{Isotropic Distribution}

The gravitational field of our system is governed by the Schwarzschild BH
and we adopt the following familiar form for the spacetime metric:
\begin{equation}
ds^2 = -(1-2M/r)dt^2 + \frac{dr^2}{1-2M/r} + r^2 d\theta^2 + r^2 {\rm sin}^2\theta d\phi^2. 
\end{equation}
The monoenergetic phase-space distribution function we shall employ everywhere in 
space is given by
\begin{equation}
\label{f}
f(E) = K \delta (E-E_{\infty}), \ \ \ K, E_{\infty} \ {\rm constant},
\end{equation}
where $E=-{\bf p \cdot e_t}$ is the ``energy at infinity" of a particle per unit mass, ${\bf p}$ is 
the particle energy-momentum 4-vector, divided by its rest mass, and ${\bf e_t} = 
\partial/\partial t$ is the time coordinate basis vector. All particles are
assumed to have the same rest mass $m$.
The constant $\Einf$ satisfies $\Einf > 1$, appropriate for an unbound orbit, while the 
constant $K$, given $\Einf$, is determined by the asymptotic density at infinity
[see Eq.~\ref{K} below].
This distribution function, specifically chosen to describe a homogeneous density of 
unbound, collisionless particles moving randomly at the same speed far from the BH, 
in fact applies everywhere as a consequence of 
Liouville's theorem. Our adopted  function, though simple, may also be thought of as a Green's
function for other energy distribution functions (e.g., Maxwellians or power laws) and our 
results below in which $\Einf$ appears can always be integrated over $\Einf$ for 
these alternative energy distributions.

\subsubsection{Density}

We now repeat the
derivation of the number density of particles at all radii $r$ we provided previously (see
Appendix A in Ref~\cite{ShaT85}), so that we may refer to and modify some of the equations in subsequent sections
where we change the boundary conditions or contrast the results with other cases.
The particle energy per unit mass measured
by a local, static, orthonormal observer with 4-velocity ${\bf u} = {\bf e_{\hat t}}$ at radius $r$ is 
given by
\begin{equation}
\label{Eloc}
E_{local} \equiv p^{\hat t} = -{\bf p \cdot e_{\hat t}} = \frac{E}{(1-2M/r)^{1/2}},
\end{equation}
where a caret on a variable denotes an orthonormal component.
The number density of particles at $r$ measured by this observer is then 
\begin{eqnarray}
\label{nisoGR}
n(r) &=& \int f(E) d^3 \phat \\
\label{n2STGR}
     &=& 4 \pi \int f(E) \phat^2 d\phat,
\end{eqnarray}
where
\begin{equation}
\label{phat}
\phat = [(p^{\hat t})^2 -1]^{1/2}
\end{equation}
is the particle 3-momentum per unit mass. Using Eqns.~(\ref{Eloc}) and ~(\ref{phat}) we obtain
\begin{equation}
\label{p3}
\phat^2 d\phat = \frac{\phat E dE}{1-2M/r} = \frac{(E^2-1+2M/r)^{1/2}}{(1-2M/r)^{3/2}} E dE. 
\end{equation}
Substituting Eqs.~(\ref{f}) and ~(\ref{p3}) into ~(\ref{n2STGR}) gives
\begin{equation}
\label{n3d}
n(r) = 4 \pi K \frac{(\Einf^2-1+2M/r)^{1/2}\Einf}{(1-2M/r)^{3/2}}.
\end{equation}
Evaluating Eq.~(\ref{n3d}) at $r=\infty$ determines $K$:
\begin{equation}
\label{K}
K = \frac{\ninf}{4 \pi  (\Einf^2-1)^{1/2}\Einf}.
\end{equation} 
So in general the density profile is
\begin{equation}
\label{n3dGR}
\frac{n(r)}{\ninf} = \frac{(\Einf^2-1+2M/r)^{1/2}}{(\Einf^2-1)^{1/2}(1-2M/r)^{3/2}}.
\end{equation}

An important special case applies to nonrelativistic particles at infinity moving with velocity 
$\vhatinf \ll 1$, for which
\begin{equation}
\label{EinfNR}
\Einf \approx 1+\frac{1}{2}\vhatinf^2, \ \ \ \Einf^2 -1 \approx 2(\Einf-1) \approx \vhatinf^2.
\end{equation}
In this limit Eq.~(\ref{K}) reduces to 
\begin{equation}
\label{Knr}
K \approx \frac{\ninf}{4 \pi \vhatinf}, \ \ \ \ (\vhatinf \ll 1),
\end{equation}
and Eq.~(\ref{n3d}) becomes
\begin{equation}
\label{n3dnr}
\frac{n(r)}{\ninf} \approx \frac{[1+2M/(r \vhatinf^2)]^{1/2}}{(1-2M/r)^{3/2}}, \ \ \ \ (\vhatinf \ll 1).
\end{equation}
%We note that Eq.~\ref{n3dnr} is consistent with the Eq.~(14.2.22) in 
%Shapiro \& Teukolsky~(\cite{ShaT83}, hereafter ST), 
%who derived the Newtonian density profile, which is applicable only when $v \ll 1$ and $M/r \ll 1$.
We note that Eq.~(\ref{n3dnr}) in the weak-field domain, wherein $M/r \ll 1$, is consistent with earlier 
Newtonian derivations in Refs.~\cite{ZelN71,ShaT83}.
%and is immediately recognized as a statement of 
%Newtonian energy conservation.

\subsubsection{Velocity Dispersion}

The velocity dispersion may be calculated formally from
\begin{equation}
\label{vdisp}
\langle \vhat^2(r) \rangle=\frac{\int \vhat^2 f(E) d^3\phat}{\int f(E) d^3 \phat},
\end{equation}
where the locally measured velocity is $\vhat = \phat/E_{local}$
or, using Eqs.~(\ref{Eloc}) and (\ref{phat}), 
\begin{equation}
\label{vhat}
\vhat = \frac{(E^2-1+2M/r)^{1/2}}{E}.
\end{equation}
Substituting Eqs.~(\ref{f}) and (\ref{vhat}) into (\ref{vdisp}) and integrating
yields
\begin{equation}
\label{v3d}
\langle \vhat^2(r) \rangle = \vhat^2(\Einf) 
                           =\frac{(\Einf^2-1+2M/r)}{\Einf^2}.
\end{equation}
The first equality in Eq.~(\ref{v3d}) also may be arrived at 
trivially by noting that all particles move on geodesics with the same specific 
energy $\Einf$, and the second equality also can be obtained by invoking
$E_{local} = \gamma = 1/(1-\vhat^2)^{1/2}$, replacing $E_{local}$ by $E$ using
Eq.~(\ref{Eloc}), and inverting for $\vhat$. 

\subsubsection{Accretion Rate}

Here we generalize our earlier Newtonian derivation of the DM accretion rate onto the BH 
[see Ref.~\cite{ShaT83}, Eqs.~(14.2.13)-(14.2.20)] to one that is fully general relativistic.
The phase-space momentum element $d^3 \phat$ may be expressed as
\begin{equation}
\label{d3p}
d^3 \phat = 2 \pi p^{\hat \bot} d p^{\hat \bot} d p^{\hat r} =
            \frac{4 \pi J dJ dE}{(1-2M/r)^{1/2} |v^{\hat r}| r^2},
\end{equation}
where $\bot$ denotes directions perpendicular to the radial direction (whereby
$J = r p^{\hat \bot}$ is the particle angular momentum per unit mass) and where  
Eqs.~(12.4.9), (12.4.13) and (12.4.16) of ~Ref.~\cite{ShaT83} were employed to
relate $dp^{\hat r}$ to $dE$: 
\begin{equation}
\label{prhat}
dp^{\hat r} = \frac{dE}{(1-2M/r)^{1/2} |v^{\hat r}|}.
\end{equation}
An additional factor of 2 arises in Eq.~(\ref{d3p}) 
since for a given $E$, $p^{\hat r}$ can be either positive or negative.
Let $N^{-}(r,E,J)$ be the number of particles per interval $dr, dE$, and $dJ$ with 
{\it inward}-directed radial
velocity:
%\begin{eqnarray}
%\label{N}
%N^{-}(r,E,J) dr dE dJ &=& \frac{1}{2} f(E,J) d^3r d^3\phat \nonumber \\ 
%                      &=& 8 \pi^2 \frac{J dJ dE dr}{(1-2M/r)^{1/2}}
%                       \frac{f(E,J)}{|v^{\hat r|}}. \ \ \ \
%\end{eqnarray}
\begin{eqnarray}
\label{N}
N^{-}(r,E,J) dr dE dJ &=& \frac{1}{2} f(E,J) d^3r d^3\phat \\
                      &=& 
                       \frac{8 \pi^2 J f(E,J)}{|v^{\hat r|}(1-2M/r)^{1/2}}
                       dr dE dJ. \nonumber
\end{eqnarray}

The resulting total capture rate for particles onto the central BH as 
measured by the local observer is then

\begin{eqnarray}
\label{dNdtau}
\frac{dN_{tot}}{d\tau} &=& \int^\infty_{(1-\frac{2M}{r})^{\frac{1}{2}}} dE 
               \int_0^{J_{min}(E)} dJ 
         |v^{\hat r}| N^{-}(r,E,J) \ \ \ \ \ \ \ \ 
               \nonumber\\
                     &=& 8 \pi^2 \int^\infty_{(1-\frac{2M}{r})^{\frac{1}{2}}} dE
               \int_0^{J_{min}(E)} dJ \frac{Jf}{(1-\frac{2M}{r})^{1/2}},
\end{eqnarray}
where particles moving with specific angular momentum less than a critical value
$J_{min}$ will be captured by BH as they approach pericenter.
The region $J < J_{min}$ thus defines a capture loss cone. For nonrelativistic
(NR) particles at large distances with $E-1 \ll 1$, which is the case
for typical stars in clusters and galaxies, and cold dark matter, we note that 
\begin{equation}
\label{JminNR}
J_{min}(E) = 4M  \ \ \ \ ({\rm NR \ particles},\ \vhatinf \ll 1).
\end{equation}
Substituting Eqs.~(\ref{f}) and ({\ref{K}) into
Eq.~(\ref{dNdtau}) and evaluating the integral yields
\begin{equation}
\label{dNdtau2}
\frac{dN_{tot}}{d\tau} = \frac{\pi \ninf J_{min}^2(\Einf)}
                         {\Einf(\Einf^2 - 1)^{1/2}}
                         \frac{1}{(1-2M/r)^{1/2}}.
\end{equation}

The particle capture rate as measured by a static observer at infinity is then
given by
\begin{equation}
\label{dNdtinf}
\frac{dN_{tot}}{dt}=\frac{dN_{tot}}{d\tau}\frac{d\tau}{dt} = 
                 \frac{dN_{tot}}{d\tau}(1-\frac{2M}{r})^{1/2},
\end{equation} 
or 
\begin{equation}
\label{dNdt}
\frac{dN_{tot}}{dt} = \frac{\pi \ninf J_{min}^2(\Einf)}
                         {\Einf(\Einf^2 - 1)^{1/2}}.
\end{equation}
It is this observer who measures the steady depletion of particles from the ambient cluster
and their acquisition by the BH. 

The corresponding {\it rest-mass} accretion rate measured by this observer, 
$dM_0/dt = m dN_{tot}/dt$, is then given by
\begin{equation}
\label{dM0dt}
\frac{dM_0}{dt} = \frac{\pi \rhoinf J_{min}^2(\Einf)}
                         {\Einf(\Einf^2 - 1)^{1/2}},
\end{equation}
where $\rhoinf \equiv m \ninf$ is the asymptotic {\it rest-mass} density.
Equation~(\ref{dM0dt}) agrees with the result obtained by a different 
route in Ref~\cite{GamGDNS21}, [see their
Eq.~(76), setting $\gamma_{\infty}=\Einf$ and $L_c(\Einf) = m J_{min}(\Einf)$].
The rate of accretion of {\it total mass-energy} onto the black hole 
is then given by $dM/dt = m \Einf dN/dt$ or
\begin{equation}
\label{dMdt}
\frac{dM}{dt} = \frac{\pi \rhoinf J_{min}^2(\Einf)}
                         {(\Einf^2 - 1)^{1/2}}.
\end{equation}

Evaluating Eqs.~(\ref{dM0dt}) and (\ref{dMdt}) 
for NR particles at infinity, using Eqs.~(\ref{EinfNR})
and ~(\ref{JminNR}),
yields
\begin{equation}
\label{dMdtNR}
\frac{dM}{dt} \approx \frac{dM_0}{dt} \approx \frac{16 \pi M^2 \rhoinf}{\vhatinf}   \ \ \ \ ({\rm NR \  particles},  \ \vhatinf \ll 1),
%\frac{dM}{dt} \approx \frac{16 \pi M^2 \rhoinf}{\vhatinf}   \ \ \ \ 
\end{equation}
which agrees with the previous Newtonian result we derived in Ref.~\cite{ShaT83}
[see Eq.~(14.2.20)]. 
Since the steady-state accretion rate can be
evaluated at large $r \gg M$, as we did in Ref.~\cite{ShaT83}, it is no surprise 
that the Newtonian derivation for the rate provided there for nonrelativistic 
particles at infinity yields the exact same
result derived in GR for the limiting case given by Eq.~(\ref{dMdtNR}).
Note that to convert back from geometric to physical units, one simply multiplies
the right-hand sides of Eqs.~(\ref{dM0dt})--(\ref{dMdtNR}) by $G^2/c^2$.

It is interesting to note that the particle density as measured by
a locally static observer {\it blows up} when the observer is stationed arbitrarily
close to the BH horizon [see Eq.~(\ref{n3dGR}) as $r \rightarrow 2M$].
(However, recall that static observers cannot exist at or inside the horizon.)
Using Eq.~(12.4.17) of Ref.~\cite{ShaT83}, for the particle radial velocity, i.e.,
\begin{equation}
\label{vr}
| v^{\hat r}| 
= \left[ 1 - \frac{1}{E^2} (1-\frac{2M}{r})(1+\frac{J^2}{r^2}) \right]^{1/2},
\end{equation} 
it is seen that the particles all have velocities
which are measured by this observer to approach the speed of light and 
move in the radial direction as they approach the horizon. 
Based on these observations one might naively take the radial matter flux to be  $\sim n(r) \vhat$, 
and then estimate the accretion rate by multiplying this product by the invariant 
area $4 \pi r^2$. Doing so yields a rate that blows up near the horizon. 
However, this estimate is too naive, as the actual steady-state depletion rate of
unbound particles from the ambient gas must be obtained by more careful
considerations, such as those we incorporated above, leading to the net accretion 
rates onto the BH as measured by a distant, static observer, 
Eqs.~(\ref{dNdt})-(\ref{dMdt}). The latter remain 
perfectly {\it finite}. This contrast is sufficiently striking 
that we are motivated to 
provide two alternative derivations in the Appendix for the net accretion rate, 
leading to the same equations obtained above.

The integration of an isotropic distribution function $f(E)$ over all phase
space allows for so-called ``white hole" orbits, i.e., outgoing trajectories at the horizon. 
Eliminating these orbits leads to a density profile of the form $n(r) \sim 1/(1-2M/r)^{1/2}$ as 
$r \rightarrow 2M$. Thus, we still have $n(r)$ blowing up near the horizon, but only
with a mild (``redshift") factor instead of the $1/(1-2M/r)^{3/2}$ behavior exhibited by
Eq.~(\ref{n3dGR}). Using this ``modified" density near the horizon to measure the
inward flux then yields the correct locally measured accretion rate, Eq.~(\ref{dNdtau2}). 
Furthermore, choosing to measure the density by an observer comoving with the net flow should lead 
to a finite density at the horizon~\cite{RioS17a,RioS17b}. 
However, as these modifications do not affect 
our computations of the accretion rates, which are our primary targets, 
we do not implement them here.

\subsubsection{Massless and Extremely Relativistic Particles} \label{photon}

To evaluate Eq.~(\ref{dMdt}) for  massless
particles (e.g. photons), we may first set $\Einf = (1-\vhatinf^2)^{-1/2}$
and then note that $J_{min} = b(\Einf^2-1)^{1/2} = \vhatinf (1-\vhatinf^2)^{-1/2}$,
where $b$ is the critical impact parameter for particle capture. 
Now $b = 3 \sqrt{3}$ is the critical impact parameter for 
massless particles [see, e.g., Ref.~\cite{ShaT83}, Eqs.~(12.4.36) 
and (12.5.11)]. Next let $\ninf m \Einf \equiv \epsilon_{\infty}^r$ be the 
energy density of particles far from the BH. Then taking the limit as 
$\vhatinf \rightarrow 1$ with these substitutions gives
\begin{equation}
\label{dMdtphoton}
\frac{dM}{dt} = 27\pi M^2 \epsilon_{\infty}^r \ \ \ (m=0),
\end{equation}
which agrees with the result quoted in Exercise 14.4 in Ref.~\cite{ShaT83} that
was obtained by an alternative approach.

We also note that Eq.~(\ref{dMdtphoton}) also applies to extremely relativistic 
(ER) particles with nonzero rest mass $m$. For such particles the accretion
rates may also be written as
%\begin{equation}
%\frac{dM_0}{dt} = 27 \pi M^2 \rhoinf   \ \ \ 
%({\rm ER \ particles}, \vhatinf \gg 1),
%\end{equation}
%and
\begin{equation}
\frac{dM}{dt} = \Einf \frac{dM_0}{dt} \approx 27 \pi M^2 \Einf \rhoinf   \ \ \ 
({\rm ER \ particles}, \vhatinf \gg 1).
\end{equation}

\subsection{Loss-Cone Effect}

Given that particles having nearly radial, inward velocities result in their
being captured by the BH, it is interesting to examine a scenario
in which, after a sufficient time has passed and steady state
is achieved, there is a depletion of particles in a low-angular momentum 
capture loss cone about the BH. For perfectly collisionless gas  
in the complete absence of perturbations,
the loss cone cannot be replenished, since particle self-interactions and
gravitational scattering by stars or other perturbers are assumed absent. (For scenarios
in which self-interactions of dark matter particles around BHs
may be important, see, e.g., Refs.~(\cite{ShaP14,Sha18,DelK23}) and
where their gravitational scattering off stars may be significant, 
see, e.g., Refs.~(\cite{Mer04,GneP04,ShaH22}), 
and references therein). In reality, the slightest gravitational perturbations, 
at large distances, due, for example, to small density anisotropies or to intruders, or the
weakest self-interactions, will likely be sufficient to 
replenish the narrow loss cone. Our  analysis above of the matter profiles and accretion rate
will thus apply in these situations. But immediately below we will consider a perturbation-free, 
perfectly  collisionless cluster that has depleted its loss cone and cannot refill it.

To describe this idealized situation  
a minimal modification to our adopted distribution function will suffice:
\begin{eqnarray}
\label{floss}
f(E) &=& K \delta(E-\Einf),  \ \  J_{min}(E) \leq J \leq J_{max}(E), \nonumber \\
     &=& 0, \ \ 0 \leq J \leq J_{min}(E),
\end{eqnarray}
where again $K$ and $\Einf > 1$ are constants.

\subsubsection{Density}

To determine the density profile we again evaluate Eq.~(\ref{nisoGR}),
employing Eq.~(\ref{d3p}), which yields
\begin{equation}
\label{nisoloss}
n(r)=\int^\infty_{(1-\frac{2M}{r})^{\frac{1}{2}}} f(E) dE
               \int_{J_{min}(E)}^{J_{max}(E)}
               \frac{4 \pi J dJ}{r^2 |v^{\hat r}| (1-\frac{2M}{r})^{1/2}}.
\end{equation} 
where $|v^{\hat r}|$ is given by Eq.~(\ref{vr}) and $J_{max}(E)$ is the
maximum specific angular momentum that a 
particle moving on a geodesic with specific energy $E$ can have,
\begin{equation}
\label{Jmax}
J_{max}(E) = r\left[\frac{(E^2 -1 +\frac{2M}{r})}{(1-\frac{2M}{r})}\right]^{1/2}
\end{equation} 
[set $dr/d\tau=0$ and $\tilde{l} = J_{max}(E)$ in Eq.~(12.4.13) 
in Ref.~\cite{ShaT83}]. Substituting Eq.~(\ref{floss}) into (\ref{nisoloss})
and integrating yields
\begin{equation}
\label{nisoloss2}
\frac{n(r)}{\ninf} = \frac{[\Einf^2-(1-2M/r)(1+
\left(J_{min}(E)/r \right)^2)]^{1/2}}
{(\Einf^2-1)^{1/2}(1-2M/r)^{3/2}},
\end{equation}
where the normalization constant $K$ is again given by Eq.~(\ref{K}). 
Comparing Eqs~(\ref{n3dGR}) and (\ref{nisoloss2}) shows that the
spike density is somewhat lower everywhere, but mostly around 
$r \lesssim  J_{min}$, if a loss cone is established and unreplenished.

Equation~(\ref{nisoloss2}) reduces to Eq.~(\ref{n3dGR}) in the absence of a loss cone,
i.e. when the captured particle distribution is assumed to be continually 
replenished so that $J_{min}(E) = 0$ in Eq.~(\ref{floss}). For nonrelativistic
particles with $\vhatinf \ll 1$ and an empty 
loss cone with $J_{min} = 4M$  the density becomes

\begin{eqnarray}
\label{nisonewt}
\frac{n(r)}{\ninf} &\approx& 
\frac{ [ 1+2M/(r\vhatinf^2)
%- (J_{min}/\vhatinf r)^2 \left( 1-2M/r \right) ]^{1/2} }
(1 -\frac{8M}{r} + \left( \frac{4M}{r} \right)^2) ]^{1/2} }
{(1-2M/r)^{3/2}},\nonumber \\
%&&     \ \ \ \ \vhatinf \ll 1,
&&     \ \ \ \ ({\rm NR \ particles},\ \vhatinf \ll 1).
\end{eqnarray}
For sufficiently small $\vhatinf \ll 1$ the density
exhibits a minimum at $r\approx 4M$  outside the event horizon.

\subsubsection{Velocity Dispersion}

As all particles have the same energy $\Einf$, they have the same
velocity and velocity dispersion profile as in the absence of a 
loss cone, i.e., Eq.~(\ref{v3d}).

\subsubsection{Accretion Rate}

Given a loss cone that is unreplenished with particles 
once they have been captured, the accretion rate becomes zero in steady state.

\section{Unbound, Monoenergetic, Razor-Thin Disk}

\subsection{Isotropic Distribution}

We now derive the surface density profile around the BH for unbound, collisionless
particles assumed to reside in a razor-thin disk. 
Since the distribution function is zero outside of the disk, 
we will simply work in the plane of the disk and drop the vertical dependence 
of any quantity. For comparison purposes we 
adopt the same  distribution function adopted for the large 3D cluster,
but now confined to the 2D disk plane, i.e.,
we again use Eq.~\ref{f} to describe unbound particles moving isotropically in the plane 
of the disk. 
\subsubsection{Density}

The surface number density $\Sigma{\text *}(r)$ measured by a locally static observer 
is then given by
\begin{equation}
\label{sig0}
\Sigma{\text *}(r) =  \int f(E) d^2 \phat = 2 \pi \int f(E) \frac{ E dE}{1-2M/r},
\end{equation} 
where we used Eq.~(\ref{p3}) to evaluate $d^2 \phat = 2 \pi \phat d\phat$.
Substituting Eq.~(\ref{f}) in (\ref{sig0}) and integrating yields
\begin{equation}
\frac{\Sigma{\text *}(r)}{\ \ \Sigma{\text *}_\infty} = 
\frac{\Sigma(r)}{\ \ \Sigma_\infty} = \frac{1}{1-2M/r},
\end{equation}
where $\Sigma(r) = m \Sigma{\text *}(r)$ is the surface rest-mass density, $m$ is the
particle rest mass and $K = \Sigma_\infty /(2 \pi m \Einf)$. Far from the
BH the surface density is thus seen to be flat. Once again, eliminating the
``white hole" orbits softens the blowup near the horizon, whereby the
modified surface density scales as $\Sigma(r) \sim 1/(1-2M/r)^{1/2}$ as 
$r \rightarrow 2M$. 

\subsubsection{Velocity Dispersion}

As all particles again have the same energy $\Einf$, they have the same
velocity and velocity dispersion profile given by  Eq.~(\ref{v3d}).

\subsubsection{Accretion Rate}

To derive the accretion rate we adapt Eq.~(\ref{N}) to a plane, yielding
\begin{eqnarray}
\label{N2d}
N^{-}(r,E,J) dr dE dJ &=& \frac{1}{2} f(E,J) d^2r d^2\phat \nonumber \\
                      &=& 4 \pi \frac{ dJ dE dr}{(1-2M/r)^{1/2}}
                       \frac{f(E,J)}{|v^{\hat r|}}, \ \ \ \
\end{eqnarray}
where we used $d^2r = 2 \pi r dr$, $J=p^{\hat \bot} r$, Eq.~(\ref{prhat}) and
\begin{equation}
\label{dp2}
d^2\phat = d p^{\hat \bot} d p^{\hat r} = 4 \frac{dJ}{r} 
                            \frac{dE}{(1-2M/r)^{1/2} |v^{\hat r}|}. 
\end{equation}
The factor of $4$ is inserted since for a given $E$ and $J$,
both $p^{\hat r}$ and $p^{\hat \bot}$ can be either
positive or negative. 

The resulting capture rate measured by the local observer is then
given by
\begin{eqnarray}
\label{dNdtaud}
\frac{dN_{tot}}{d\tau} &=& \int^\infty_{(1-\frac{2M}{r})^{\frac{1}{2}}} dE
               \int_0^{J_{min}(E)} dJ
               |v^{\hat r}| N^{-}(r,E,J) \ \ \ \ \ \ \ \
               \nonumber\\
                     &=& 4 \pi \int^\infty_{(1-\frac{2M}{r})^{\frac{1}{2}}} dE
               \int_0^{J_{min}(E)} dJ \frac{f}{(1-\frac{2M}{r})^{1/2}},
\end{eqnarray}
yielding
\begin{equation}
\frac{dN_{tot}}{d\tau} = \frac{2 \Sigma{\text *}_{\infty} J_{min}(\Einf)}{\Einf}
                         \frac{1}{(1-2M/r)^{1/2}}.
\end{equation}
The depletion rate of the disk as measured by a distant static observer is
then obtained using Eq.~(\ref{dNdtinf}), which gives
\begin{equation}
\frac{dN_{tot}}{dt} = \frac{2 \Sigma{\text *}_{\infty} J_{min}(\Einf)}{\Einf}.
\end{equation}
The corresponding rates of accretion of rest mass and total mass-energy
onto the BH are then
\begin{equation}
\label{dM0dtdisk}
\frac{dM_0}{dt}= \frac{2  \Sigma_{\infty} J_{min}(\Einf)}{\Einf}
\end{equation}
and
\begin{equation}
\label{dMdtdisk}
\frac{dM}{dt}= 2  \Sigma_{\infty} J_{min}(\Einf),
\end{equation}
respectively.
The rate for NR particles at infinity, using Eq.~(\ref{JminNR}), is
then
\begin{equation}
\label{dMdtNRdisk}
dM/dt \approx 8 M \Sigma_{\infty}   \ \ ({\rm NR \ particles},\ \vhatinf \ll 1).
\end{equation}
We note that Eq.~(\ref{dMdtNRdisk}) agrees with Eq.~(93) in Ref.~\cite{CieMO22} for
Maxwell-J\"uttner particle temperatures approaching zero and Schwarzschild BHs 
[set $\alpha \equiv a/M =0$ and note 
$\rho_{s,\infty} \equiv \Sigma_{\infty}$ in Eq.~(93)]. 

It is interesting to observe that in this slow-velocity limit, the above rate of
mass-energy accretion for a 2D razor-thin disk depends only on the
surface density of distant particles and not on their velocity dispersion,
in contrast to accretion from a large 3D cluster, which depends on both
the asymptotic density and the velocity dispersion [see Eq.~(\ref{dMdtNR})].

We note that to convert back from geometric to physical units,
one simply multiplies the right-hand sides of 
Eqs.~(\ref{dM0dtdisk})-(\ref{dMdtNRdisk}) by $G^2/c$.

\subsubsection{Massless and Extremely Relativistic Particles}

Defining the asymptotic particle surface energy density to be 
$\mathcal{E}_{\infty}^r \equiv \Sigma_{\infty} \Einf$ and noting that
$J_{min} \approx 3\sqrt{3} M \Einf$ when $\Einf \gg 1$, we can
evaluate Eq.~(\ref{dMdtdisk}) for massless particles,
yielding
\begin{equation}
\label{photondisk}
\frac{dM}{dt} = 6 \sqrt{3} M \mathcal{E}^r_{\infty},  \ \ \ (m=0).
\end{equation}
We again note that Eq.~(\ref{photondisk}) also applies to ER particles
with nonzero $m$. For such particles the accretion rates may also be
written as
\begin{equation}
\label{dMdtERdisk}
\frac{dM}{dt} = \Einf \frac{dM_0}{dt} \approx 6\sqrt{3} M \Sigma_{\infty} \Einf
 \ \ \ ({\rm ER \ particles}, \vhatinf \gg 1).
\end{equation} 
We point out that Eq.~(\ref{dMdtERdisk} agrees with Eqs.~(124) and (125)
in Ref.~\cite{CieMO22} for
Maxwell-J\"uttner particle temperatures approaching infinity and 
Schwarzschild BHs [again set $\alpha \equiv a/M =0$ and note 
$\rho_{s,\infty} \equiv \Sigma_{\infty}$ in Eq.~(124) and $\epsilon_{s,\infty}
= \Sigma_{\infty} \Einf$ in Eq.~(125)].

\subsection{Loss-Cone Effect}

\subsubsection{Density}

Here we treat the scenario whereby collisionless particles in the loss cone are never
replenished once captured, whereby the distribution function may again be
represented by Eq.~(\ref{floss}). Using Eq.~(\ref{dp2}), 
the surface density is then given by
\begin{eqnarray}
\label{sigloss}
\Sigma{\text *}(r) &=&  \int f(E) d^2 \phat  \\
     &=& \int^\infty_{(1-\frac{2M}{r})^{\frac{1}{2}}} f(E) dE
               \int_{J_{min}(E)}^{J_{max}(E)}
               \frac{4 dJ}{r |v^{\hat r}| (1-\frac{2M}{r})^{1/2}} \nonumber
\end{eqnarray}
Substituting Eqs.~(\ref{floss}) and (\ref{Jmax}) and integrating yields
\begin{equation}
\label{sigloss2}
\frac{\Sigma(r)}{\Sigma_{\infty}} = \frac{ 
 1 - \frac{2}{\pi} \arctan\left( \frac{J_{min}/r}
           {\left[ \frac{\Einf^2 - 1 +2M/r}{1-2M/r} 
           -(J_{min}/r)^2)   \right]^{1/2}}    \right)  
                                     }{1-2M/r}.
\end{equation}
For NR particles in Newtonian gravitation, 
Eq.~(\ref{sigloss2}) reduces to
\begin{equation}
%%\frac{\Sigma}{\Sigma_{\infty}} \approx 1 -  
%%                     \frac{8M}{\pi r \vhatinf}, 
%%               \ \ (M/r \ll \vhatinf \ll 1).
\frac{\Sigma(r)}{\Sigma_{\infty}} \approx 1 - \frac{2}{\pi} 
                     \arcsin\left( \frac{4M}{r \vhatinf} \right),\\ 
%               \ \ (NR \ {\rm particles}, \ \vhatinf \ll 1, \ M/r \ll 1).
               \ \ (\vhatinf \ll 1, \ M/r \ll 1).
\end{equation}

\subsubsection{Velocity Dispersion}

As all particles again have the same energy $\Einf$, they have the same
velocity and velocity dispersion profile given by  Eq.~(\ref{v3d}).

\subsubsection{Accretion Rate}

Given a loss cone that is unreplenished with particles
once they have been captured, the accretion rate becomes zero in steady state.

\section{Comparison of Accretion Rates}

Comparing the accretion rates of 3D clusters and 2D thin-disks with filled
loss cones is not entirely
straightforward, given their different geometries and defining parameters.
The results will depend on the different physical systems that 
are compared. To give one example, let us consider a large, homogeneous 
spherical cluster of particles with density $\rhoinf$, 
isotropic velocity dispersion $\vinf$, radius $R_c \gg r_h$, and total mass $M_c$. 
Imagine that it undergoes
collapse parallel to the $z$-axis to a thin pancake in the $x-y$ plane,
preserving its surface density along cylinders. This scenario might mimic
one way that thin sheets of collisionless particles form in the
early universe. We will compare the unbound,
collisionless particle accretion rates for the large spherical cluster 
and the pancake for NR particles with $\vinf \ll 1$.

The surface density in the pancake is
\begin{equation}
\Sigma_{\infty}(r_{\perp}) = 2 \rhoinf R_c \left(1 - \frac{r^2_{\perp}}{R^2_c} \right)^{1/2},
\end{equation}
where $r_{\perp}$ is the radius in the $x$-$y$ plane measured from the center
of the pancake.
Taking the surface density at $r_{\perp}$ in the central core, requiring 
$r_h \ll r_{\perp} \ll R_c$, 
whereby the particles remain largely unperturbed by the central BH, gives
\begin{equation}
\label{sigpan}
\Sigma_{\infty} \approx 2 \rhoinf R_c.
\end{equation}
The ratio of BH accretion rates in the spherical cluster {\it vs.} the thin disk is
then approximated by
\begin{equation}
\label{ratiovd1}
\frac{\dot M_c}{\dot M_d} \approx \pi \frac{M}{R_c} \frac{1}{\vinf},
\end{equation}
where $\dot M_c$ is given by Eq.~(\ref{dMdtNR}) and $\dot M_d$ is given 
by Eq.~(\ref{dMdtNRdisk}),
substituting Eq.~(\ref{sigpan}) for $\Sigma_{\infty}$. We can evaluate
$\vinf$ if we assume
that it is some fraction of the virial value,
whereby
\begin{equation}
\label{virial}
\vinf^2 \approx \frac{M_c}{R_c}.
\end{equation}
Equation~(\ref{ratiovd1}) then yields
\begin{equation}
\label{ratiovd2}
\frac{\dot M_c}{\dot M_d} \approx \pi \left( \frac{M}{M_c} \right)^{1/2}
                       \left( \frac{M}{R_c} \right)^{1/2} \ll 1.
\end{equation}
The strong inequality above for this one extreme example shows 
the dominance of disk {\it vs.} cluster accretion, thereby demonstrating 
that {\it the rate of unbound, collisionless particles 
can depend significantly on the geometry of the ambient particle cloud,
other parameters being equal}.  

It is also interesting to compare the collisionless particle rates
to the standard spherical Bondi accretion rate for particles comprising 
a true fluid.  The Bondi rate for an adiabatic gas with 
$1 \leq \Gamma \leq 5/3$ ~\cite{Bon52} (see also~ \cite{ShaT83} for a relativistic
treatment) is given by
\begin{equation}
\label{dMdtbondi}
\frac{dM}{dt} = \dot M_b = 4 \pi \lambda \frac{ M^2 \rhoinf}{a^3_{\infty}},
\end{equation}
where $a_{\infty}$ is the asymptotic sound speed of the fluid and $\lambda$ 
is a constant of order unity that depends on $\Gamma$.
We will equate $a_{\infty}$ to $\vinf$ for comparison below, obtaining
\begin{equation}
\label{ratiobv}
\frac{\dot M_b}{\dot M_c} \approx \frac{\lambda}{4} \frac{1}{\vinf^2} \gg 1,
\end{equation} 
and
\begin{equation}
\label{ratiobd}
\frac{\dot M_b}{\dot M_d} \approx \frac{\pi \lambda}{4} \frac{M}{R_c}
            \frac{1}{\vinf^3} \approx  
            \frac{\pi \lambda}{4} \left(\frac{M}{M_c} \right)
            \left( \frac{R_c}{M_c} \right)^{1/2},
\end{equation}
where we used Eq.~(\ref{virial}) to obtain the second 
equality in Eq.~(\ref{ratiobd}).
Equation~(\ref{ratiobv}) shows that Bondi accretion dominates collisionless
particle accretion for a large 3D cluster. However
Eq.~(\ref{ratiobd}) suggests that the ratio for an extended 2D thin disk 
depends on the particular system, since the first factor in parentheses 
on the right-hand side may be much smaller than unity, but the second factor is 
much bigger than unity. Once again, spatial geometry counts.

\subsection{Applications}

There may exist 
a near ``universal" value of the surface density for DM that spans, within a factor of two,
over at least nine (and possibly more) galaxy magnitudes and across
several different Hubble types~\cite{DGSFWGGKW09,KorF04}:
\begin{equation}
\Sigma_{\infty}^u \approx 140~{\rm M_{\odot}~pc^{-2}}.
\end{equation}
The cosmological implications of this observation are not yet resolved, but 
Eq.~(\ref{dMdtNRdisk}) suggests that a thin disk with the surface density 
$\Sigma_{\infty}^u$ and 
a central BH would have an accretion rate given by
\begin{equation}
\label{dMdtu}
\frac{dM^u_d}{dt} \approx 3.3 \times 10^{-2} {\rm M_{\odot}~yr^{-1}} 
                    \left( \frac{M}{10^6 {\rm M_{\odot}}} \right)
                    \left( \frac{\Sigma_{\infty}^u}
                        {140~{\rm M_{\odot}~yr^{-1}}} \right).
\end{equation}

The surface density $\Sigma_{\infty}^u$ is also within a factor of two
of estimates of DM in the Galactic neighborhood,
where $\rho_D \sim 0.008~{\rm M_{\odot}~pc^{-3}}$ and $D \sim 8.5~{\rm kpc}$
~\cite{BovT12}, yielding $\Sigma_{\infty}^{Gal} \sim \rho_D D 
\approx 68~{\rm M_{\odot}~pc^{-2}}$.
Since $\rho_D$ may scale as $r^{-1}$ should it obey an NFW profile 
~\cite{NavFW97}, the Galactic
DM surface density estimated as $\sim \rho_D r$ would be constant all the way down
to the BH sphere of influence at $r_h$. If it were to reside in a thin disk near 
the Galactic Center its accretion rate would be comparable to the universal value 
given by Eq.~(\ref{dMdtu}) for a BH of mass 
$4.3 \times 10^6$~\cite{GenEG10,GheSWLD08}. If instead it were to
occupy a large spherical cluster outside $r_h$ and move with a velocity
dispersion of $\sim 100~{\rm km~s^{-1}}$ then Eq.~(\ref{dMdtNR}) suggests it will accrete
at a much smaller rate of 
$dM^c/dt \approx 2.4 \times 10^{-7}~{\rm M_{\odot}~yr^{-1}}$.

By comparison the Bondi value for the baryon accretion rate onto Sgr A* at the Galactic
Center, which is determined from the gas density and temperature inferred 
from the diffuse X-ray emission observed by {\it Chandra} at 
$\sim 2$ arcsec ($\sim 0.1$ pc) from the black hole, is 
$dM^b/dt \sim 2 \times 10^{-5}~{\rm M_{\odot}~yr^{-1}}$.
In fact, the baryon accretion rate is believed to be $\sim 10^{-8}~{\rm M_{\odot}~yr^{-1}}$,
or roughly three orders of magnitude below the Bondi value as determined 
from polarization measurements~\cite{MarMZR07} and models of the near-horizon
accretion flow and emitted luminosity~\cite{ShcB10,ResTQG17}. 
This difference may be due to the angular momentum
of the stellar winds that may be supplying the gas, or possibly to more
exotic effects such as the heating of the gas by DM annihilation in the spike about
the BH~\cite{BenGBS19}.

\section{Summary and Conclusions}

We have examined the steady-state density and velocity profiles, and the
associated accretion rates, of collisionless particles
(e.g. stars or DM) moving around a central Schwarzschild black hole in unbound orbits.
We considered two distinct spatial geometries for the particle: an infinite
3D cluster and a 2D razor-thin disk, both without net angular momentum. 
We adopted the same simple  monoenergetic, phase-space 
distribution function for the particles for both cases, arguing that, though idealized, 
this assignment was sufficient to illustrate the features that might distinguish 
nonrotating spherical-like and
disklike collisionless systems orbiting a black hole. We treated both a totally
isotropic velocity profile at each point and one in which an empty loss cone
is present due to the capture of low-angular momentum particles that are captured by 
the BH and not replenished. In all cases the net angular momentum of the systems was
assumed to be zero so that the only differences were due to the different spatial
geometries and velocity anisotropies adopted.

We found that even in the weak-field region, where $r \gg M$, a mild spike arises in the locally 
measured particle density $n(r)$  for the 3D cluster 
but that the surface density $\Sigma^*$ remains constant with $r$ for the 2D razor-thin disk.
We also found that, at least for one simple application, the rate of accretion of the
disk was much larger than that of the cluster. However, both rates were much lower than
the Bondi accretion rate for a fluid with a comparable particle density and velocity
dispersion (i.e., sound speed) far from the BH.

While these differences may not be so striking when more realistic phase-space 
distribution functions and geometries are considered, the results do suggest that 
the spatial distribution of particles around a black hole is a feature that affects the
resulting steady-state particle profiles and accretion rates significantly. So this is just one
other factor that must be accounted for, in addition to knowing what the nature of the
particles are (e.g., collisionless or collisional fluid matter, or mildly collisional 
by virtue of self-interactions and/or annihilations) 
and the global properties of their
distributions (bound or unbound, with or without net angular momentum, subject or not to
gravitational intruders, etc) in assessing their profiles and capture rates about a black hole.

\medskip

{\it Acknowledgments}: It is a pleasure to thank Yuk Tung Liu and Olivier Sarbach for 
several useful discussions. This paper was supported in part by NSF Grants No. 
PHY-2006066
and No. PHY-2308242 and NASA Grant 80NSSC17K0070 to the University of Illinois at
Urbana-Champaign.

\appendix

\section{Cluster Accretion Rates: Alternative Derivations}\label{A1}

\subsection{Alternative Derivation 1}

Here we obtain the accretion rate for an unbound, monoenergetic, nonrotating cluster 
as measured by a locally static observer by
calculating the total inward particle flux across a sphere of radius $r$ 
$\times$ the area of the sphere $\times$ the fraction of these particles that
move within the loss cone and are thus captured: 
\begin{equation}
\label{altdNdtau}
\frac{dN_{tot}}{d\tau} = \left( \frac{1}{4} n \vhat \right) \times
                       \left( 4 \pi r^2 \right) \times
                       \cal{P},
\end{equation}
where $\vhat$ is the magnitude of the isotropic 3-velocity at $r$ given
by Eq.~(\ref{v3d}) and
$\cal{P}$ is the fraction of particles captured. 
%Combining $\phat = \gamma = 1/(1-\vhat^2)^{1/2}$,
%with Eqs.~(\ref{Eloc}) and (\ref{phat}) yields
%\begin{equation}
%\vhat = \frac{(\Einf^2 - 1 + 2M/r)^{1/2}}{\Einf},
%\end{equation} 
%where we used the fact that our adopted distribution function 
%endows all particles with the identical specific energy, $E=\Einf$.
For $\cal{P}$ we have
\begin{equation}
{\cal{P}} = \frac{\int_0^{J_{min}(\Einf)} dJ J}{\int_0^{J_{max}(\Einf)} dJ J}
        = \frac{J_{min}^2(\Einf)}{J_{max}^2(\Einf)},
\end{equation}
while the density $n$ is given by Eq.~(\ref{n3dGR}) and $J_{max}(E)$ is given
by Eq.~(\ref{Jmax}).
Assembling the factors in Eq.~(\ref{altdNdtau}) then yields Eq.~(\ref{dNdtau2})
for the locally measured accretion rate at $r$, $dN_{tot}/d\tau$, from which, 
using Eq.~(\ref{dNdtinf}), the 
rates measured by a distant observer, Eqs.~(\ref{dNdt}) - (\ref{dMdt}) 
for $dN_{tot}/dt$, $dM_0/dt$ and $dM/dt$, respectively, follow immediately.

\subsection{Alternative Derivation 2}

Here we provide yet another derivation of the accretion rate found above.
The maximum impact parameter $b_{max}$ for a particle of energy $\Einf$ 
falling inward from infinity to be captured by the BH is given by
\begin{equation}
b_{max}^2 = \frac{J_{min}^2(\Einf)}{\Einf^2 -1},
\end{equation}
[see, e.g. Ref.~\cite{ShaT83}, Eq.~(12.4.35)], with the typo corrected for the 
missing square on $\Einf$), whereby the capture cross section is 
\begin{equation}
\label{sigma}
\sigma_{cap} = \pi b_{max}^2 = \frac{\pi J_{min}^2(\Einf)}{\Einf^2 -1}.
\end{equation}
So far from the BH the accretion rate is obtained as
the intensity of particles for an isotropic distribution 
$\times$ the area of a large sphere about the BH $\times$ the solid angle
within which a particle is captured:
\begin{equation}
\label{dNdt2}
\frac{dN_{tot}}{dt} = \left( \frac{\ninf \vhatinf}{4 \pi} \right) (4 \pi r^2) 
                      (\Delta \Omega_{cap}),
\end{equation}
where using Eq.~(\ref{v3d}) at $r \rightarrow \infty$, we have
\begin{equation}
\vhatinf = \frac{(\Einf^2-1)^{1/2}}{\Einf}
\end{equation}
and where
\begin{equation}
\label{omega} 
\Delta \Omega_{cap} = \frac{\sigma_{cap}}{r^2}.
\end{equation}
Assembling the factors in Eq.~(\ref{dNdt2}) again yields 
Eq.~(\ref{dNdt}) for $dN_{tot}/dt$, from which 
Eq.~(\ref{dNdtau2}) for $dN_{tot}/d\tau$ and Eqs.~(\ref{dM0dt}) - (\ref{dMdt})
for $dM_0/dt$ and $dM/dt$, respectively, again follow immediately.

\bibliography{paper}
\end{document}